# Particle-Hole Symmetry Breaking in the Pseudogap State of Bi2201

M. Hashimoto*[1, 2, 3], R.-H. He*[1, 2], K. Tanaka[1, 2, 3, 4], J. P. Testaud[1, 2, 3], W. Meevasana[1, 2], R. G. Moore[1, 2], D. H. Lu[1, 2], H. Yao[1], Y. Yoshida[5], H. Eisaki[5], T. P. Devereaux[1, 2], Z. Hussain[3], & Z.-X. Shen[1, 2]

[1]*Stanford Institute for Materials and Energy Sciences, SLAC National Accelerator Laboratory, 2575 Sand Hill Road, Menlo Park, CA 94025*

[2]*Geballe Laboratory for Advanced Materials, Departments of Physics and Applied Physics, Stanford University, CA 94305*

[3]*Advanced Light Source, Lawrence Berkeley National Lab, Berkeley, CA 94720, USA*

[4]*Department of Physics,* Osaka University, Toyonaka, Osaka 560-0043, Japan

[5]*Nanoelectronics Research Institute, AIST, Ibaraki 305-8568, Japan*

*These authors contributed equally to this work.

**In conventional superconductors, a gap exists in the energy absorption spectrum only below the transition temperature ($T_c$), corresponding to the energy price to pay for breaking a Cooper pair of electrons. In high-$T_c$ cuprate superconductors above $T_c$, an energy gap called the pseudogap exists, and is controversially attributed either to pre-formed superconducting pairs, which would exhibit**

**particle-hole symmetry, or to competing phases which would typically break it [1]. Scanning tunnelling microscopy (STM) studies suggest that the pseudogap stems from lattice translational symmetry breaking [2-9] and is associated with a different characteristic spectrum for adding or removing electrons (particle-hole asymmetry) [2,3]. However, no signature of either spatial or energy symmetry breaking of the pseudogap has previously been observed by angle-resolved photoemission spectroscopy (ARPES) [9-18]. Here we report ARPES data from Bi2201 which reveals both particle-hole symmetry breaking and dramatic spectral broadening indicative of spatial symmetry breaking without long range order, upon crossing through $T^*$ into the pseudogap state. This symmetry breaking is found in the dominant region of the momentum space for the pseudogap, around the so-called anti-node near the Brillouin zone boundary. Our finding supports the STM conclusion that the pseudogap state is a broken-symmetry state that is distinct from homogeneous superconductivity.**

The nature of the pseudogap can be explored by examining the dispersion of the occupied electronic states measured by ARPES. As shown in Fig. 1m, when a particle-hole symmetric gap opens from the normal state dispersion (red curve) due to homogeneous superconductivity, one always expects an alignment between Fermi momentum $k_F$ and the "back-bending" or saturation momentum (green arrows) of the dispersion in the gapped states (weighted blue curve). Because of this strong constraint, the observation of back-bending or dispersion saturation anomaly away from $k_F$ in a gapped state can be a conclusive evidence of a particle-hole symmetry broken nature of the gap, even though the information of the unoccupied state may be missing. Here we note that the dispersion of the spectral peak position, regardless of the spectral weight, is the simple and direct way to address the issue of particle-hole symmetry. Due to suppressed weight that makes it hard to discern from experimental background, the back-bending may sometimes be subtle and may show up as a dispersion saturation. Nethertheless, it can be distinguished from a smooth dispersion where no dispersion saturation anomaly occurs.

Fig. 1 shows the temperature evolution across $T^*$ of the Fermi-Dirac function (FD) divided ARPES spectra of a nearly optimally-doped cuprate superconductor $Pb_{0.55}Bi_{1.5}Sr_{1.6}La_{0.4}CuO_{6+\delta}$ (Pb-Bi2201) with $T_c \sim 34$ K. Data were taken in an antinodal cut approximately along $(\pi, -\pi)$-$(\pi, 0)$-$(\pi, \pi)$ as shown in the inset of Fig. 1g. The spectra in the true normal state above $T^*$ (~125 K) present a parabolic dispersion of the intensity maximum as a function of momentum with two clear Fermi level crossings at $k_F$'s (red spectra) and a bottom reaching $E_{Bot} \sim -20$ meV at $(\pi, 0)$ (Fig. 1a,1g & 1n). Given that the data were taken at 160K, the measured spectra in the true normal state are remarkably simple, similar to that of ordinary metal [19]. Whereas this is naturally expected for a band in the absence of a gap, with the pseudogap opening around $E_F$ below $T^*$, the spectra become surprisingly incoherent and the spectral weight centroid is transferred towards higher binding energy (Fig. 1a-l). Despite the broadness of the spectra at low temperatures, the intensity maximum of each spectrum can be easily defined and traced as a function of momentum. The dispersion thus extracted becomes stronger towards lower temperatures with the band bottom at $(\pi, 0)$ being pushed far away from the true normal state $E_{Bot}$ as summarized in Fig. 1n. Well below $T^*$ where the spectra are fully gapped in the measured antinodal cut, no dispersion saturation or back-bending is observed at $k_F$, defined by the Fermi crossing at $T = 160$ K (guides to the eyes in red). Instead, while approaching $E_F$, the dispersion flattens and appears to bend back at momenta (green arrows) markedly away from $k_F$. This misalignment can be also seen in the summary of data in Fig. 1i-1l (the red spectra are for $k_F$ and the green spectra are for possible back-bending momenta). Contrasting to what is expected in Fig. 1m, the behaviour below $T^*$ is completely different from the expected dispersion in homogeneous superconducting state, suggesting that the transition from the true normal state above $T^*$ to the pseudogap state has a different origin. In addition to the misalignment between $k_F$ and the back-bending momenta for $T \ll T^*$, the observed widening of the separation between $E_{Bot}$ and energy at the bend-back momenta upon lowering temperature also contradicts the expectation in Fig. 1m where the energy separation (or the overall dispersion decreases with an opening of a superconducting gap.

The connection between the pseudogap formation and the dispersion effects may be directly examined from the temperature dependence of spectra seen at two representative momenta, $(\pi, 0)$ and $k_F$ (Fig. 2a and 2b). With increasing temperature, the energy position of the spectral intensity maxima continuously evolve towards and ultimately reach $E_{Bot}$ and $E_F$, respectively, at $T^* \sim 125\pm10$ K (Fig. 2c) where the pseudogap closes, consistent with temperature from other reports of different measurements on Bi2201 near optimal doping [9, 10, 20, 21]. The smooth temperature evolution upon cooling down from the true normal state suggests that the dispersion of the intensity maximum found at $T \ll T^*$ (Fig. 1n) is directly related to the pseudogap physics. That is, the misalignment between $k_F$ and the momenta where back-bending of the dispersion occur at $T \ll T^*$ is due to the pseudogap opening. This suggests that the pseudogap state violates the momentum structure expected from a particle-hole symmetric homogeneous superconducting state. This finding goes beyond the earlier STM work that suggests the lattice symmetric breaking nature of the low temperature phase, but lacks the detailed temperature dependence linking the symmetry breaking and pseudogap opening.

The link of the observed spectral change and pseudogap formation can also be found in the temperature dependence of the spectral intensity itself. In Fig. 2d we plot spectral intensity at $k_F$ in the energy window of [-0.36, -0.30] eV as a function of temperature, with the spectra normalized either by the photon flux, which reflect raw spectral weight, or by a selected energy window as indicated in the figure (See Supplementary Fig. 3 for the spectra with these normalizations). Regardless of the normalization used, once again, $T^*$ is the temperature below which a strong temperature dependence emerges. The data further suggest that a very high energy scale is involved in the spectral weight redistribution associated with pseudogap formation. In addition, as energy-integrated spectral weight is conserved at a fixed momentum [14], a change in spectral intensity over a wide occupied energy window at $k_F$ could involve spectral weight redistribution from unoccupied states, which is often connected to particle-hole symmetry breaking. This is consistent with the particle-hole symmetry breaking upon the pseudogap opening observed in the momentum structure (Fig. 1n). All above observations can be robustly reproduced and no sample degeneration was found during the measurement (Supplementary Figs. 4-6).

The opening of a pseudogap below $T^*$ that breaks the particle-hole symmetry is accompanied with several anomalous temperature-dependent behaviours. When considering the spectral lineshape evolution with temperature based on a simple textbook picture for a metal, one expects a broadening of the spectra with increasing temperature due to the increased phase space available for inelastic electron scattering. Further, the opening of a superconducting energy gap at low temperatures generally sharpens quasi-particles by suppressing density of states for inelastic electron scattering. In stark contrast, the observed spectral evolution below $T^*$ goes to the opposite direction, suggesting an unusual broadening mechanism that sets in the pseudogap state. In addition, the spectra no longer retain a Lorentzian line shape below $T^*$. Our data also provide important information missed in earlier studies. The pseudogap has previously been reported to disappear in a subtle manner, manifested as a filling-in of the gapped states in the vicinity of $E_F$ (a few $k_B T$) rather than a clear gap closing [9-11]. Different from those previous reports, our results uncover the pseudogap closing with a clearly shifting energy scale (Fig. 2c) and a concurrent spectral weight redistribution over a wide energy range (Fig. 2d).

Our observations collectively reveal a rich physical picture of the pseudogap phenomena beyond a simple extension of superconductivity which necessitates other ingredients for a complete picture. This picture is overall consistent and aligned with Ref. [10], but differs in critical areas such as the assumption of particle-hole symmetry near the antinode. Our finding of a particle-hole symmetry breaking supports the idea that the pseudogap state competes with superconductivity. Motivated by the observations [2-19, 22-25] as well as theoretical implications [26] of spatial symmetry breaking (density-wave-like correlations), we explore whether a density wave picture can account for the observation. We first consider a conventional model [27] of long-range density-wave orders in the weak-coupling limit. Fig.3 shows the expected ARPES spectra from a calculation with incommensurate checkerboard density-wave order of orthogonal wave vectors, [0.26$\pi$, 0] & [0, 0.26$\pi$] (3a) and the commensurate [$\pi$, $\pi$] density-wave order (3b) for three different order parameters V = 15, 30 and 60 meV (weighted blue curves from top to bottom). We find that the modelling can qualitatively reproduce some critical aspects of our findings. Most importantly, there is no dispersion anomaly at $k_F$ because the gap by these density-wave correlations opens by band-folding in a particle-hole asymmetric manner, which is very different from the gap opening by superconductivity. Rather, there exists back-bending (arrows) markedly away from $k_F$ (red dashed lines). Besides, these models provide a qualitatively better account for the dramatic shifting energy scale with temperature than the existence of homogeneous superconductivity. The most striking is the downward shift of the band bottom energy (note the similarity between 1n and 3a) which hardly happens for superconductivity. The modelling, on the other hand, fails to describe the counterintuitive broadening of spectra with decreasing temperature observed in the present study. A model assuming a short-range nature (10 lattice constants) of the [π, π] density-wave order [28, 29] can do a better job regarding the broader spectra in the gapped state (Fig. 3c). In this model, interestingly, shorter density-wave correlation length is required to reproduce the broader spectra at lower temperature. However, it is difficult to explain the wide energy range of spectral weight distribution in the pseudogap state given the relatively low $T^*$ with this model. This suggests that the textbook starting point assuming extended Bloch waves becomes no longer accurate [8]. An unconventional explanation that captures the good aspects of the simple models of competing order but with emphasis on effects of strong coupling thus seems to be necessary for our observations of the pseudogap. The highly localized nature indicated by the broadness of the spectra is consistent with this idea.

Reminiscent of the ARPES effects observed here, a particular form of spatial and particle-hole symmetry broken spectra is observed by momentum-integrated STM in the pseudogap phase at low temperature[2, 3, 30]. The observed nanoscale inhomogeneity associated with local density-wave order and dominated by high-energy states away from the $E_F$ is formally consistent with our spatially-averaged ARPES observation, both pointing to the symmetry-broken nature and high-energy relevance of the pseudogap phase. The momentum and energy resolution of ARPES allows us to directly reveal that not only the spectral weight, as found in STM experiments, but also the gap form itself has a particle-hole symmetry broken nature in the pseudogap state. Furthermore, the detailed temperature dependent data made possible by ARPES reveal: i) the strikingly simple electronic structure of the true normal state above $T^*$ - a critical baseline to understand the pseudogap state that has not been experimentally established; and ii) a direct connection to the symmetry broken state below $T^*$ with the opening of the pseudogap. As such, our finding allows an integrated picture to advance the understanding of the pseudogap as a strong coupling form of broken-symmetry state that emerges from a simple normal state above $T^*$ and most likely competes with superconductivity.

**Acknowledgements** We thank W.-S. Lee, E. Berg, K. K. Gomes, B. Moritz, S. A. Kivelson, M. Grilli, H. Q. Lin, N. Nagaosa, A. Fujimori and J. Zaanen for helpful discussions and Y. Li for experimental assistance on SQUID measurements. R.-H.H. thanks the SGF for financial support. This work is supported by the Department of Energy, Office of Basic Energy Science under contract DE-AC02-76SF00515.

**Author contributions** ARPES measurements were done by M.H., R.-H.H., K.T. and W.M. Y.Y. and H.E. grew and prepared the samples. M.H. and R.-H.H. analyzed the ARPES data and wrote the paper with suggestions and comments by J.P.T., H.Y, T.P.D. and Z.-X. S. ARPES simulations were done by M.H., R.-H.H., J.P.T., H.Y. and T.P.D. R.G.M. and D.H.L. maintained the ARPES endstation. Z.H. and Z.-X.S. are responsible for project direction, planning and infrastructure.

**Author Information** The authors declare no competing financial interests. Correspondence and requests for materials should be addressed to Z.-X.S (zxshen@stanford.edu).

**Figure 1 Particle-hole symmetry breaking in the antinodal dispersion of pseudogapped Pb-Bi2201.** $T_c$ = 34 K, $T^*$ = 125 ± 10 K. **a-l,** Fermi-Dirac function (FD) divided image plots (upper panels) and corresponding spectra as a function of parallel momentum (lower panels) taken along the antinodal cut shown in the inset of **g** at selected temperatures. The intensity maximum of each spectrum is marked by circle. Typical error bars estimated in analysis of the smoothed first-derivative spectra are smaller than the symbol size. An additional shoulder feature of weak dispersion is seen close to $E_F$ at 10 K. Note that this feature at low energy has been a major focus in previous studies [9-18]. Spectra in red and green are at $k_F$ and possible back-bending momenta of the intensity maximum dispersion, respectively (Supplementary Figs. 1 and 2). **m,** Simulated dispersion for *d*-wave homogeneous superconductivity with order parameter V = 30 meV. The quasiparticle energy for a given momentum state *k* is given by $E(k) = \sqrt{\varepsilon(k)^2 + \Delta(k)^2}$. Cuts are along (π, -π)-(π, 0)-(π, π). The red (blue) curve is for the true normal (gapped) state. Spectral weight is indicated by the curve thickness. The back-bending (or saturation) of the dispersion and $k_F$ are indicated in the panel. Note that the back-bending momentum in the gapped state remains aligned with $k_F$. See details in Supplementary Method. **n,** Summary of the intensity maximum dispersions at different temperatures. Typical error bars from the derivative analysis are smaller than the symbol size.

**Figure 2 Temperature evolution of the antinodal spectra tied to the pseudogap opening. a, b,** FD-divided spectra at (π, 0) and $k_F$, respectively, with constant vertical offset between each spectrum after the normalization at the highest binding energy, taken at different temperatures as color-coded in increments of 10 K. The intensity maxima at 160 K and 10 K are marked by circles. Note that the low-energy shoulder feature loses its clear definition roughly above $T_c$ upon raising temperature. The offset and normalization are for better visualization of the spectral line shape and the intensity maximum feature but not for the discussion of the spectral weight redistribution (See Supplementary Fig. 3 for photon flux normalization). **c,** Temperature-dependent energy position of the intensity maximum at ($\pi$, 0) and two $k_F$ along the antinodal cut. Error bars are estimated in analysis of the smoothed first-derivative spectra. **d,** Temperature dependence of the average intensity within [-0.36, -0.30] eV after spectral weight normalization by the photon flux ($I_0$) and within selected energy ranges as denoted, of the raw and FD-divided spectra at two $k_F$ (see Supplementary Fig. 3 for the normalized spectra). The average intensity is normalized at 160 K for comparison with its typical error bars (± 5% for $I_0$ normalization and smaller than the symbol size for the area normalization). Note that any finite temperature dependence indicates spectral weight redistribution beyond the window.

**Figure 3 Simulated spectra for simple options of the density-wave correlations behind the pseudogap.** Cuts are along ($\pi$, -$\pi$)-($\pi$, 0)-($\pi$, $\pi$) for all case. Two long-range orders: **a**, incommensurate checkerboard density-wave order of orthogonal wave vectors, [0.26$\pi$, 0] & [0, 0.26$\pi$] and **b**, commensurate [$\pi$, $\pi$] density-wave order. The green dashed curves are shadow bands due to the corresponding order which interact with the bare band (red) producing the renormalized band dispersions (blue) presumably observable in the ordered or locally-ordered state. Vertical red dotted lines are eye guides for $k_F$. Selected renormalized band dispersions for the same eigenstate with order parameter V = 15, 30 and 60 meV (from top to bottom), independent of $k$, are shown for each case to exemplify the misalignment between $k_F$ and the back-bending (or saddle point) momentum (arrow). Spectral weight is indicated by the curve thickness. Note that only first-order band folding are considered for simplicity in simulations for **a**. See Supplementary Fig. 7 for complete results of all long-range density-wave simulations. **c**, The [$\pi$, $\pi$] density wave order with finite correlation length: 10 lattice constants, and density wave gap $\Delta_k$ = 60 meV, independent of $k$. The extracted dispersion is shown in blue curve with its thickness proportional to the peak intensity. For **a** and **b**, note that a pronounced upshift of the band bottom commonly requires a decreasing V which inevitably results in corresponding changes of the back-bending position, in contrast to the "pinning" of it possibly observed below ~ 50 K (Fig. 1n). This could be consistent with the reported temperature independence of back-bending momenta across $T_c$ [12], and poses a challenge for the exclusive density-wave picture put in such simple ways. See details in Supplementary Method.

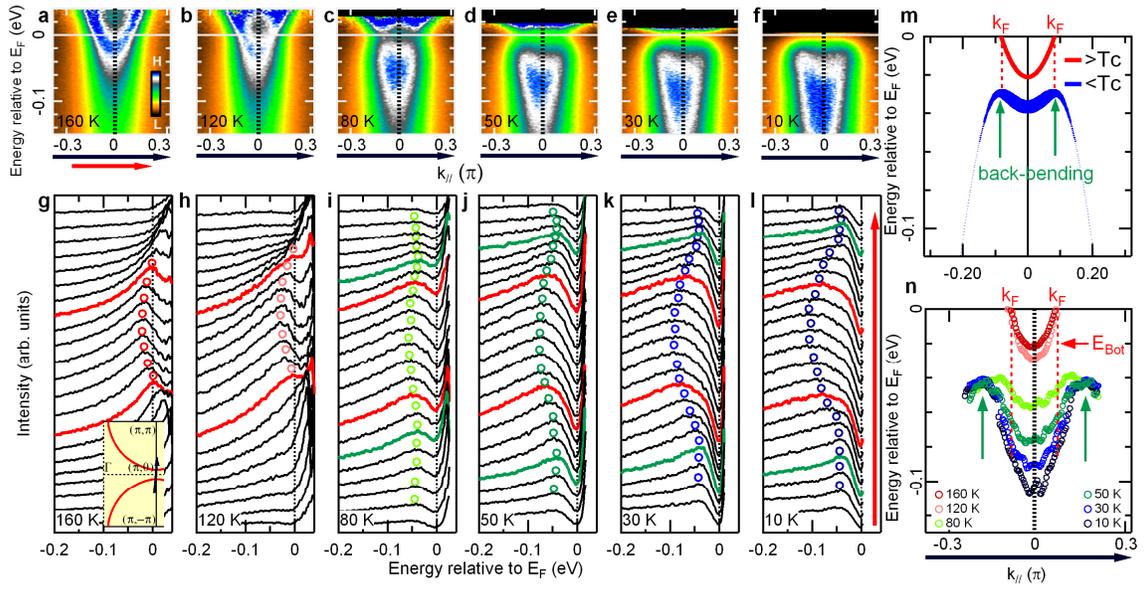

Fig. 1

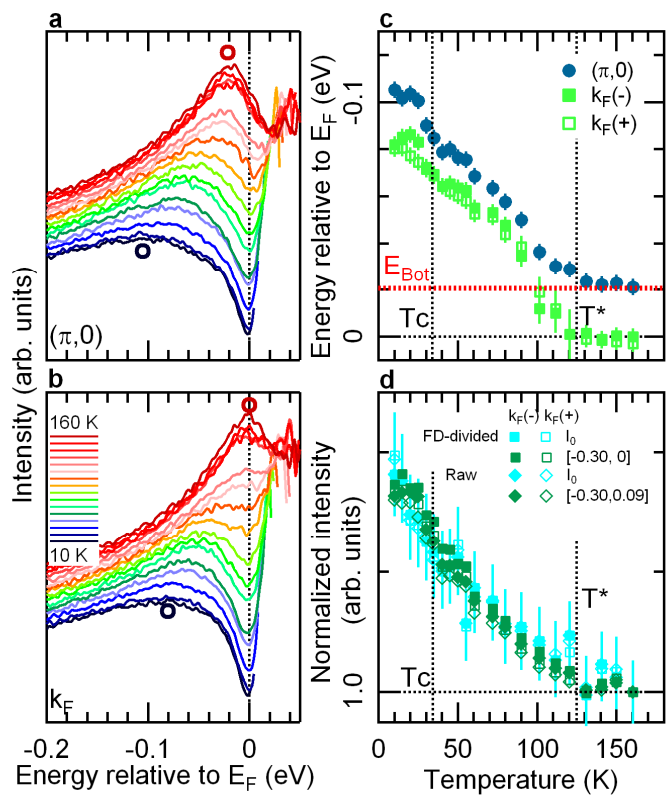

Fig. 2

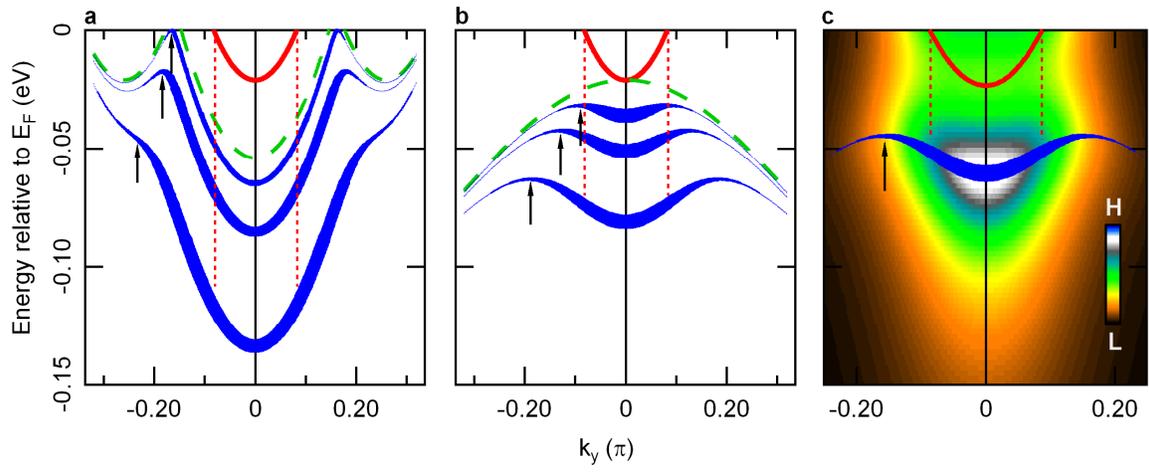

Fig. 3

# Supplementary Information for

# Particle-Hole Symmetry Breaking in the Pseudogap State of Bi2201

M. Hashimoto*[1, 2, 3], R.-H. He*[1, 2], K. Tanaka[1, 2, 3, 4], J. P. Testaud[1, 2, 3], W. Meevasana[1, 2], R. G. Moore[1, 2], D. H. Lu[1, 2], H. Yao[1], Y. Yoshida[5], H. Eisaki[5], T. P. Devereaux[1, 2], Z. Hussain[3], & Z.-X. Shen[1, 2]

*These authors contributed equally to this work.

## Samples and Experimental Method

### Materials

Nearly optimally-doped high-quality single crystals of $Pb_{0.55}Bi_{1.5}Sr_{1.6}La_{0.4}CuO_{6+\delta}$ (Pb-Bi2201) were grown by the travelling solvent floating-zone method. The Pb doping suppresses the super-modulation in the BiO plane and minimizes complications in the electronic structure due to photoelectron diffraction. This aspect, together with the absence of bilayer splitting, allows a clean tracking of the spectral evolution of pseudogapped antinodal states. The carrier concentrations of the samples were carefully adjusted by a post-annealing procedure under flowing $N_2$ gas which varies the oxygen content. All data shown were obtained on samples at the same doping level from the same batch of growth and similar post-annealings. The estimated hole concentration from the Fermi surface area is ~25 % which is consistent with previous reports for near optimal doping [31, 32]. The onset temperature of superconducting transition, $T_c$, determined by SQUID magnetometry, was 34 K with a transition width less than 3 K. X-ray and Laue diffraction showed no trace of impurity phases. The moderate size of $T^*$ (125 K ± 10 K) allows reference data to be obtained in the true normal state without involving severe thermal smearing or causing the degradation of cleaved sample surfaces (Figs. S4-S6), which would become a problem for lower dopings with higher $T^*$ and thus requiring higher measurement temperatures. Compared with higher $T_c$ cuprates, the sufficiently large difference between $T^*$ and $T_c$ benefits a systematic study of the pseudogap physics in a wide temperature range without entanglement of coherent superconductivity.

**Methods**

ARPES measurements were performed at beamline 5-4 of the Stanford Synchrotron Radiation Laboratory (SSRL) with a SCIENTA R4000 electron analyzer. All data shown in this paper were taken using 22.7 eV photons in the first (mainly) and second (partially) Brillouin zones with total energy and angular (momentum) resolutions of ~ 10 meV and ~ 0.25° (~ 0.0096 Å$^{-1}$), respectively. The temperatures were recorded closest to the sample surface position within an accuracy ±2 K. The samples were cleaved in situ at various temperatures ranging from 10 K to 160 K and measured in an ultra high vacuum chamber with a base pressure of better than 3×10$^{-11}$ Torr which was maintained below 5×10$^{-11}$ Torr during the temperature cycling. Measurements on each sample were completed within 48 hours after cleaving.

To remove the effect due to cutoff by the Fermi-Dirac (FD) function, raw ARPES spectra (e.g., Fig. S3c and d) have been divided by a convolution of the FD function at the given temperature and a Gaussian representing energy resolution. This procedure allows us to recover the actual band dispersions closest to $E_F$ and trace them above $E_F$, where thermal population leads to appreciable spectral weight especially at high temperatures. Its effect on the spectral line shape is limited to ~ $4k_BT$ around $E_F$ and our analysis of the dispersion of the maximum in spectral intensity at high energy remains robust (Fig. S2). We avoid discussing the intensity above $E_F$ where the resolution effects become stronger [13]. We used this method for the energy-distribution-curve (EDC) analysis rather than an $E_F$ symmetrisation procedure [9-11, 16, 17, 33] because the latter implicitly assumes the particle-hole symmetry of states which is shown to be broken at $k_F$ in this study. By going above $T^*$, this analysis allows us to objectively determine $k_F$, which was often assumed to be the dispersion back-bending momentum in the gapped state (so called minimum gap locus) [12]. As we show, the discrepancy between these two momenta remains finite within a wide temperature range below $T^*$.

Note here that the spectra in Fig.S3 were normalized by two methods, photon flux which gives the raw spectral weigh, or normalization within a certain energy window. The second method is used as a supplement to double check whether spectral weight is conserved within a certain energy window. If spectral weight in not conserved within the normalization window, then one should see a change outside this window of normalization. This provides a supplementary mean to check the energy scale of spectral weight redistribution.

We describe in Main Figs. 3 the modification of the band structure due to the existence of some density wave order of a wave vector $q$ based on two simple approaches, with one assuming a long-range order [27, 34] and the other taking into account short-range fluctuation effects [28, 29]. For the first case, the mean-field Hamiltonian is given by:
$H = \sum_k \varepsilon_k c_k^+ c_k + \sum_{k,q} V_q (c_{k+q}^+ c_k + h.c.)$ where $V_q$ is the interaction (order parameter) between the main band at k and the shadow band at k+$q$ generally as a result of coupling of electron to some bosonic mode (e.g., phonons for a charge density wave) of a wave vector $q$, $c_k^+$ ($c_k$) is the creation (annihilation) operators for electrons at $k$. $\varepsilon_k$ is the tight-binding bare band dispersion which is obtained by a global fit to the experimental low-energy band dispersions at 160 K. In case of the commensurate $q = q_{AF} = [\pi, \pi]$, the eigenstate $|\psi_k\rangle = u_k|k\rangle + u_{k+q_{AF}}|k+q_{AF}\rangle$ for the Hamiltonian with the eigenenergy $\varepsilon'(k)$ can be obtained by solving the matrix $M_{AF} = \begin{bmatrix} \varepsilon_k & V \\ V & \varepsilon_{k+q_{AF}} \end{bmatrix}$. In Main Fig. 3a and b and Fig. S7, the blue curve shows the renormalized band dispersion $\varepsilon'(k)$ with its thickness proportional to $|u_k|^2$. In the incommensurate checkerboard case with $q_1 = [0.26\pi, 0]$ and

$q_2 = [0, 0.26\pi]$, $M_{CB} = \begin{bmatrix} \varepsilon_{k-q_2} & 0 & V & 0 & 0 \\ 0 & \varepsilon_{k-q_1} & V & 0 & 0 \\ V & V & \varepsilon_k & V & V \\ 0 & 0 & V & \varepsilon_{k+q_1} & 0 \\ 0 & 0 & V & 0 & \varepsilon_{k+q_2} \end{bmatrix}$. Here we truncate the interaction at the first order and only take into account the first-order shadow bands in order to show a simple and clear physical picture. The qualitative existence of dispersive states at high energy and dispersion back-bendings (or saddle points) apart from $k_F$ remains robust in different calculations which apply different schemes for the truncation. Complications due to the weak but non-vanishing influence by higher orders, $nq_1$ and $nq_2$ can produce more folded bands of different spectral weight at a given momentum that can in principle comprise a broad spectral envelope reminiscent of the experimental observation. However, we note that the weak high-order constituents might be practically undetectable [34] without an anomalous suppression of the main band. The superconductivity case shown in Fig. 1m is based on the Hamiltonian $H = \sum_k \varepsilon_k c_k^+ c_k + \sum_k \Delta_k (c_{k\uparrow}^+ c_{-k\downarrow}^+ + h.c.)$, $M_{SC} = \begin{bmatrix} \varepsilon_k & \Delta_k \\ \Delta_k & -\varepsilon_{-k} \end{bmatrix}$, where a $d$-wave superconducting gap is assumed $\Delta_k = \Delta(\cos k_x - \cos k_y)/2$, with the eigenstate $|\psi_k\rangle = u_k|k\rangle + v_k|-k\rangle$, where $u_k$ and $v_k$ are the BCS coherence factors.

We consider density wave fluctuations in the spirit of Refs. 28 and 29. Specifically, we assume a spectral function $A(k,\omega) = -\frac{1}{\pi}\operatorname{Im} G(k,\omega)$, where the Green's function obeys a Dyson expression $G^{-1}(k,\omega) = G_0^{-1}(k,\omega) - \frac{1}{N}\sum_q P(q)\Delta_{k-q}^2 G_0(k-q,\omega)$, with $G_0^{-1}$ the non-interacting Green's function and $P(q)$ representing a Lorentzian distribution of ordering wave vectors having a width associated with the finite correlation length. By choosing an ordering wave vector of [π, π], the resulting spectral function is shown in Main Fig. 3c.

We note, for simplicity, we chose not to include the cases of pair density wave. Such options should also be explored in a more comprehensive theory.

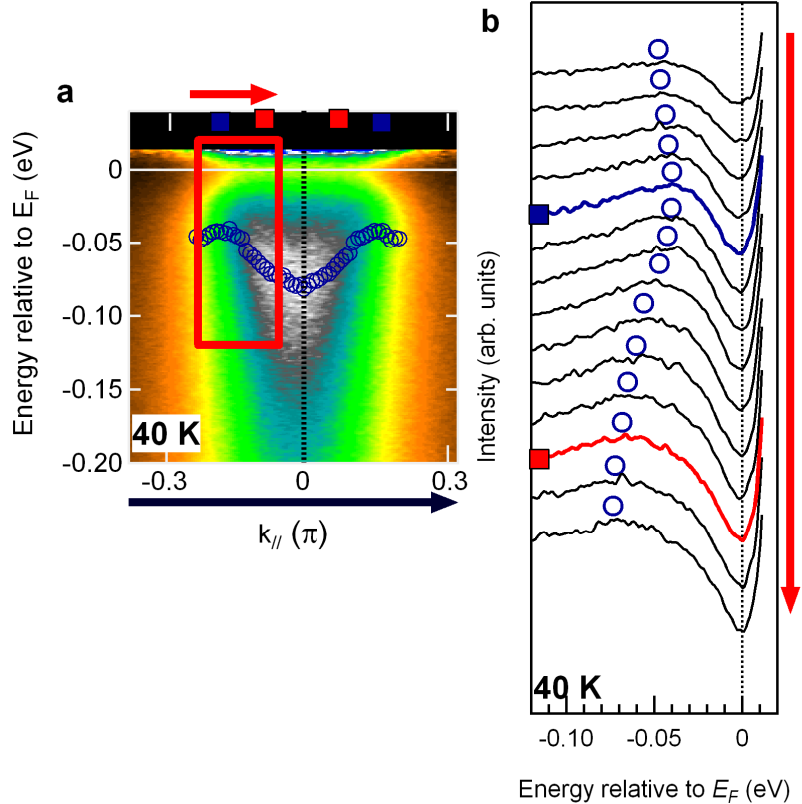

**Figure S1 Possible back-bending of the intensity maximum at 40 K. a**, Image plot of the FD-divided ARPES spectra. Red and blue squares denote the momenta of $k_F$ at 160 K and those of possible back-bendings at 40 K, respectively. **b**, EDCs in the red square region as indicated in **a**. The color EDCs correspond to the momenta as indicated by the squares. The intensity maxima of EDC are marked by blue circles, same as shown in **a**, showing the dispersion flatters at low energy where the back-bending position is defined. To estimate the intensity maxima, we first applied first derivative to the FD-divided spectra and found peak of the derivative spectra after applying moderate smoothing to eliminate the noise. We defined the maxima from the extrapolation of the smoothing-degree dependence of the peak position to zero smoothing.

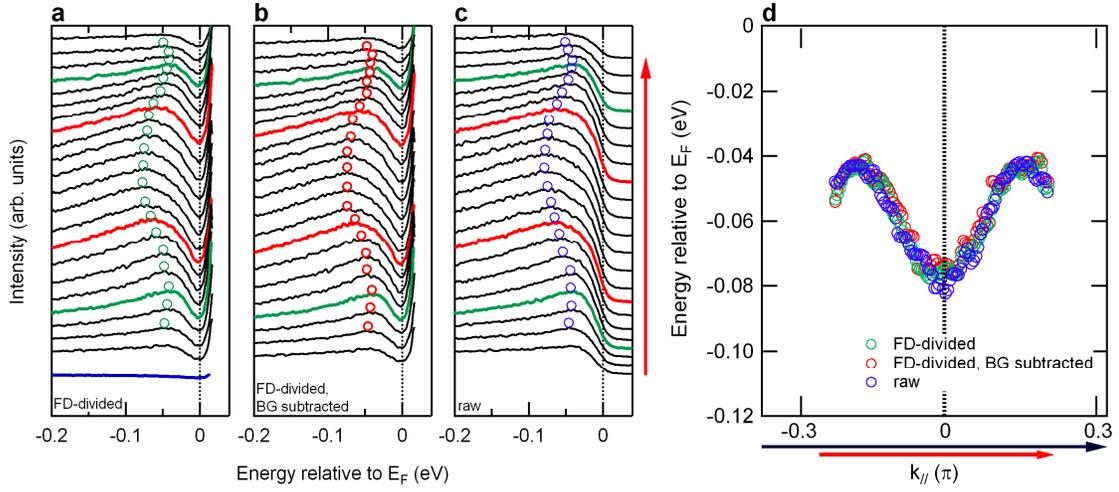

**Figure S2** Robustness of the dispersion of intensity maximum against the FD-division and background subtraction. The same data set at 50 K is used as in Main Fig. 1j. **a**, FD-divided spectra. **b**, Background (BG) subtracted FD-divided spectra. **c**, Raw spectra. Intensity maximum positions determined from the smoothed first-derivative analysis are plotted as circles in **a**-**c**. Spectra in red and green are at $k_F$ and possible back-bending momenta of the intensity maximum dispersion, respectively. Blue spectrum in **a** is a FD-divided spectrum at far from $k_F$ in the same cut, which is used for the background subtraction. **d**, Summary of the dispersions. Note that the difference between different analyses is within error bars.

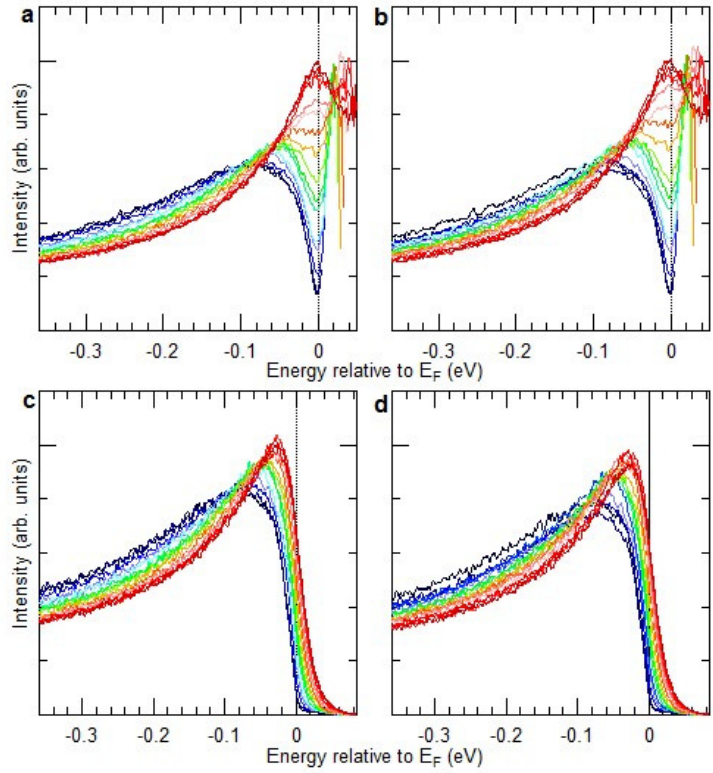

**Figure S3 Normalization dependence of the temperature evolution of FD-divided and raw EDCs at $k_F$. a,b** are for the FD-divided spectra and **c,d** for the raw spectra taken at different temperatures as color coded. The same set of data as shown in Main Fig. 2b is used. The energy windows for normalization of **a** and **c** are [-0.30, 0] and [-0.30, 0.09] eV, respectively. **b** and **d** are for the normalization by experimental photon flux. The average intensity within [-0.36, -0.30] eV is plotted in Main Fig. 2d as a function of temperature.

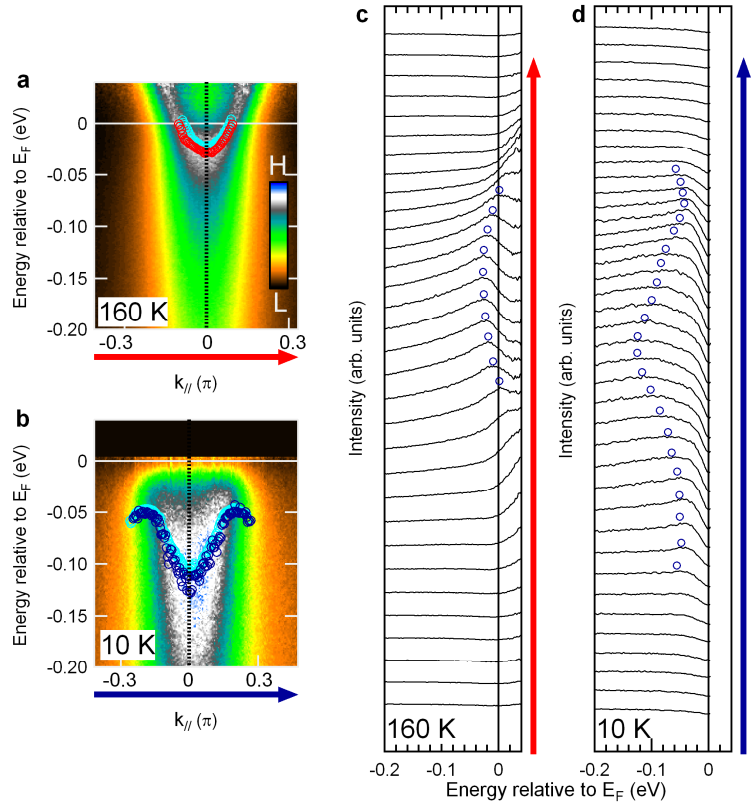

**Figure S4 Reproducibility of the antinodal spectra on different samples with three different sample cleavings at the two selected temperatures. a, b,** Image plots of the FD-divided spectra approximately along (π, -π)–(π, 0)–(π, π) at 160 K and 10 K with two different flesh cleavings at 160 K and 10 K, respectively. **c, d,** Selected EDCs from **a** and **b**, respectively. The intensity maximum at 160 K and 10 K are marked by red and blue circles, respectively. Their traceable dispersions along the cut are shown correspondingly in **a** and **b**. The dispersions obtained from another cleaving whose data are shown in the main text are overlaid (light blue circles) in **a** and **b** for comparison. Note that the cut locations are close but not exactly the same for different data sets. The shoulder features near $E_F$ (Main Fig. 1l) are also seen in **d**.

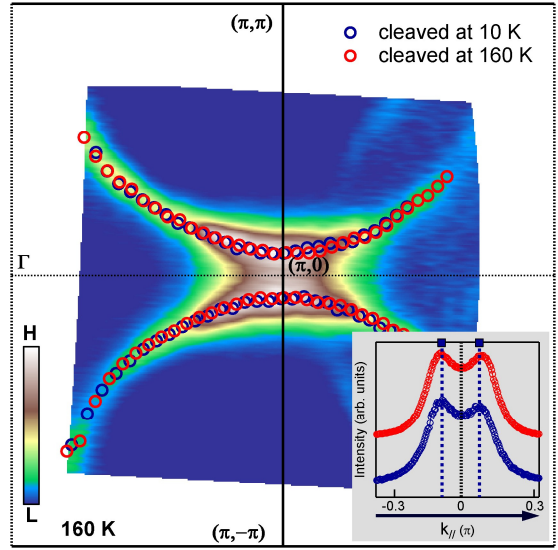

**Figure S5 Reproducibility of the ungapped Fermi surface shape at 160 K on two samples with different sample cleaving temperatures.** Data shown in the main text were taken on the sample cleaved at 10 K. The Fermi surface map at 160 K was obtained after completing the temperature dependence study from 10 K to 160 K. The spectral weight contour ($E_F \pm 10$ meV) at 160 K on another sample fleshly cleaved at 160 K with identical experimental conditions is shown here. $k_F$ determined from the momentum distribution curve (MDC) at $E_F$ for the two data sets are overlaid for comparison. The inset shows the MDCs at $E_F$ approximately along ($\pi$, -$\pi$)–($\pi$, 0)–($\pi$, $\pi$) for the two data sets. Blue squares and dashed lines are guides to eyes showing $k_F$ for the cleaving at 10 K. The shape of Fermi surface is representative of samples from the same batch at the same doping level, suggesting that no sample aging was involved in the temperature dependence study.

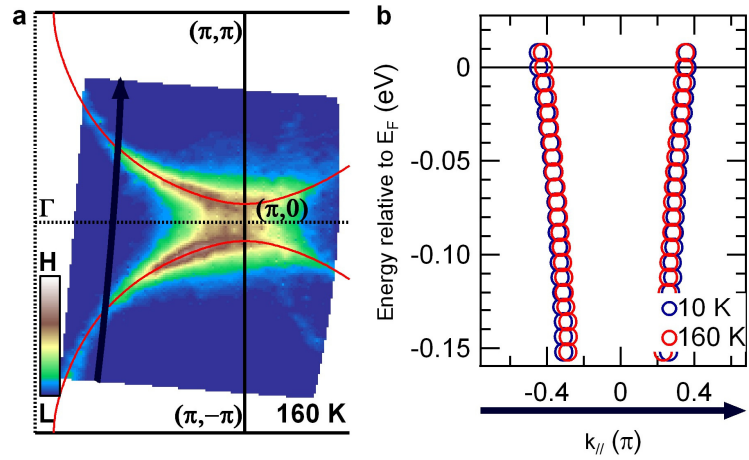

**Figure S6 Nodal dispersions without the pseudogap complication before and after the same temperature dependence study presented in the main text.** MDC dispersions along the cut shown in **a** at 10 K (blue) and 160 K (red) are plotted together in **b**, showing no detectable chemical potential shift or sample aging.

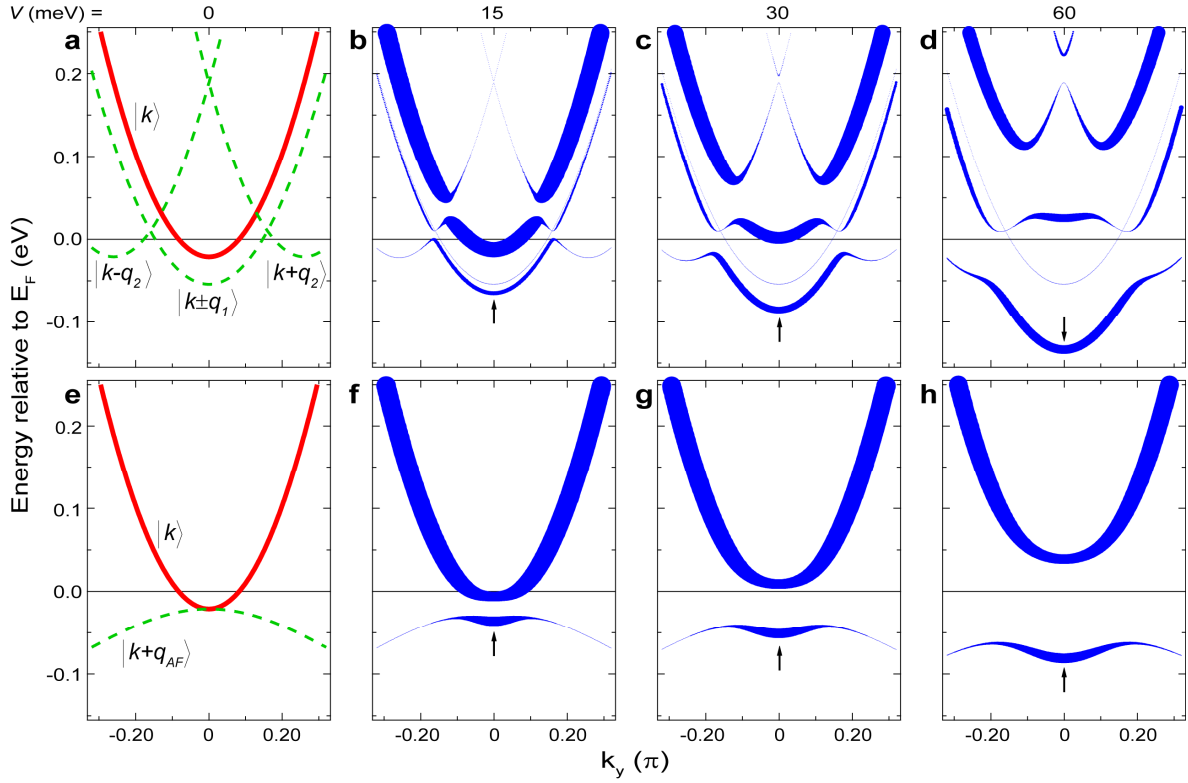

**Figure S7 Complete results of the long-range density-wave modelling. a-d** and **e-h** are for Main Fig. 3a and 3b, respectively. **a** and **e** show bare and shadow bands as specified by the corresponding eigenvector, which mutually interact to form renormalized band dispersions shown in **b-d** and **f-h** for different V. Arrows point out bands of the same eigenstates after interaction that are shown in Main Fig. 3a and 3c.

## Supplementary References